\documentclass[useAMS,usenatbib]{mn2e}
\usepackage{graphicx}

\title{HD 210111: a new $\lambda$ Bootis type SB system}
\author[Paunzen, Heiter, Fraga \& Pintado]
   {E.~Paunzen$^{1,2}$\thanks{ernst.paunzen@univie.ac.at}, U.~Heiter$^{3}$, 
   L.~Fraga$^{4}$, O.~Pintado$^{5}$\\
  $^{1}$Rozhen National Astronomical Observatory, Institute of Astronomy of the Bulgarian 
  Academy of Sciences, P.O. Box 136, \\ BG-4700 Smolyan, Bulgaria \\
  $^{2}$Institut f{\"u}r Astronomie der Universit{\"a}t Wien, 
  T{\"u}rkenschanzstr. 17, A-1180 Wien, Austria \\
  $^{3}$Department of Physics and Astronomy, Uppsala University, Box 516, SE-75120 Uppsala, Sweden \\
  $^{4}$Southern Observatory for Astrophysical Research, Casilla 603, La Serena, Chile \\
  $^{5}$Instituto Superior de Correlaci{\'o}n Geol{\'o}gica, Av. Per{\'o}n S/N, Yerba Buena, 4000 
  Tucum{\'a}n, Argentina}
\date{Accepted 2011 October 13. Received 2011 October 5; in original form 2011 April 8}

\pagerange{\pageref{firstpage}--\pageref{lastpage}} \pubyear{2012}

\def\LaTeX{L\kern-.36em\raise.3ex\hbox{a}\kern-.15em
    T\kern-.1667em\lower.7ex\hbox{E}\kern-.125emX}

\begin{document}

\label{firstpage}

\maketitle

\begin{abstract}
The small group of $\lambda$ Bootis stars comprises late B to early F-type
stars, with moderate to extreme (up to a factor 100) surface underabundances
of most Fe-peak elements and solar abundances of lighter elements (C, N, O,
and S). The main mechanisms responsible for this phenomenon are atmospheric diffusion,
meridional mixing and accretion of material from their surroundings. 
Especially spectroscopic binary (SB) systems with $\lambda$ Bootis type components are
very important to investigate the evolutionary status and accretion process in more details.
For HD~210111, also $\delta$ Scuti type pulsation was found which gives the opportunity 
to use the tools of asteroseismology for further investigations. The latter could result in
strict constraints for the amount of diffusion for this star. Together with models for the
accretion and its source this provides a unique opportunity to shed more light on these
important processes.
We present classification and high resolution spectra for HD~210111. A detailed investigation of
the most likely combinations of single star components was performed. For this, composite spectra
with different stellar astrophysical parameters were calculated and compared to the observations to
find the best fitting combination. HD~210111 comprises two equal (within the estimated errors) stars with
$T_\mathrm{eff}$\,=\,7400\,K, log\,$g$\,=\,3.8\,dex, [M/H]\,=\,$-$1.0\,dex and 
$v \sin i$\,=\,30\,km\,s$^{-1}$. This result is in line with other strict observational
facts published so far for this object. It is only the third detailed investigated $\lambda$ Bootis
type SB system, but the first one with a known IR-excess.
\end{abstract}


\section{Introduction} \label{intro}

The group of $\lambda$ Bootis stars comprise of true Population I, late B to early F-type
stars, with moderate to extreme (up to a factor 100) surface underabundances
of most Fe-peak elements and solar abundances of lighter elements (C, N, O,
and S). Only a maximum of about 2\% of all objects in the relevant spectral domain are found
to be $\lambda$ Bootis type stars (\citealt{Pau02}).

\citet{Mic86} suggested that the peculiar chemical 
abundances on the stellar surfaces are due to accretion of circumstellar material that 
is mixed in the shallow convection zone of the star by the joint action of gravitational 
settling and radiative acceleration. It naturally explains why the anomalous abundance 
pattern is similar to that found in the gas phase of the interstellar medium (ISM), 
where refractory elements like iron and silicon have condensed into dust grains. 

\citet{Kam02} and \citet{Mar09} 
developed a model which is based on the interaction of the star with its local 
ISM environment. Different levels of underabundance are produced by different 
amounts of accreted material relative to the photospheric mass. The small fraction of 
this star group is explained by the low probability of a star-cloud interaction and 
by the effects of meridional circulation, which washes out any accretion pattern a 
few million years after the accretion has stopped. The hot end of this model is 
due to strong stellar winds for stars with $T_\mathrm{eff}$\,$>$\,12\,000\,K whereas 
the cool end at about 6500\,K is defined by convection which prevents the accreted 
material to manifest at the stellar surface.

Up to now, there are at least eight double-lined spectroscopic binary (SB2) systems with 
suspected $\lambda$ Bootis candidates known (\citealt{Pau02}).
A detailed abundance analysis was done only for HD\,84948 and 
HD\,171948 so far (\citealt{Hei02a} and \citealt{Ili02}). Both SB2 systems consist of true
$\lambda$ Bootis type objects. The analysis of such SB2 systems is very important
for the above described model because it shows that both components of the system are $\lambda$ Bootis type stars.
It is clearly in contradiction with the suggested scenario by 
\citep{Far04} who suggested that two unresolved solar abundant stars mimic
a metal-weak single star spectrum. The limitations and the inconsistencies 
of this scenario with
accurate photometric measurements was already discussed by \citet{Stu06}.

In this paper we present the detection of HD~210111 (HR~8437, HIP~109306) being a
true $\lambda$ Bootis type SB system. New high resolution and classification spectroscopy
is analysed and discussed in the context of already published data. This object, also
showing $\delta$ Scuti type pulsation, is
especially interesting because an IR excess was already detected which makes it an
ideal test case for the accretion model.

\section{Observations} \label{obs}

\begin{figure}
\begin{center}
\includegraphics[width=85mm]{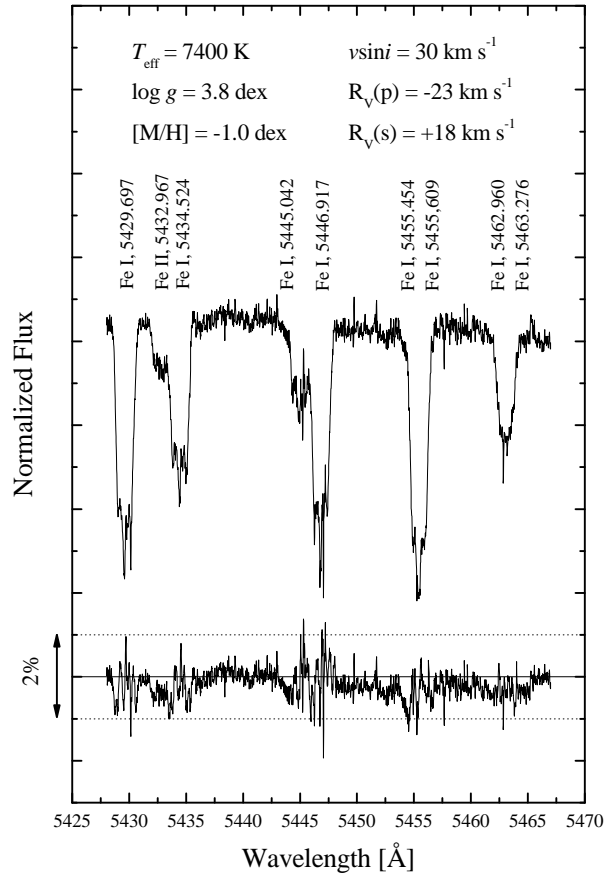}
\caption[]{Upper curve: A part of the averaged UVES spectrum. Lower curve: The difference 
between the best fit synthetic composite spectrum from a wide variety of single star 
spectrum combinations and the observed spectrum.}
\label{hrs}
\end{center}
\end{figure}

The high resolution spectra, used for the abundance and stellar parameter
estimation (Fig. \ref{hrs}), were taken from the UVES Paranal
Observatory Project (\citealt{Bag03}). In total, 24 spectra from the night of
07.07.2002 are available
which were observed with a slit width of 0.5\AA, resulting in 
a spectral resolution of about 80000. The final, averaged, spectrum was normalized
using standard IRAF routines\footnote{Available from http://iraf.noao.edu/}. 

We compared the line profiles of the averaged spectrum to those of the
individual spectra to exclude a misinterpretation due to an incorrect
merging of the data. The line profiles due to the SB2 nature are clearly
visible in all spectra.

\begin{figure}
\begin{center}
\includegraphics[width=85mm]{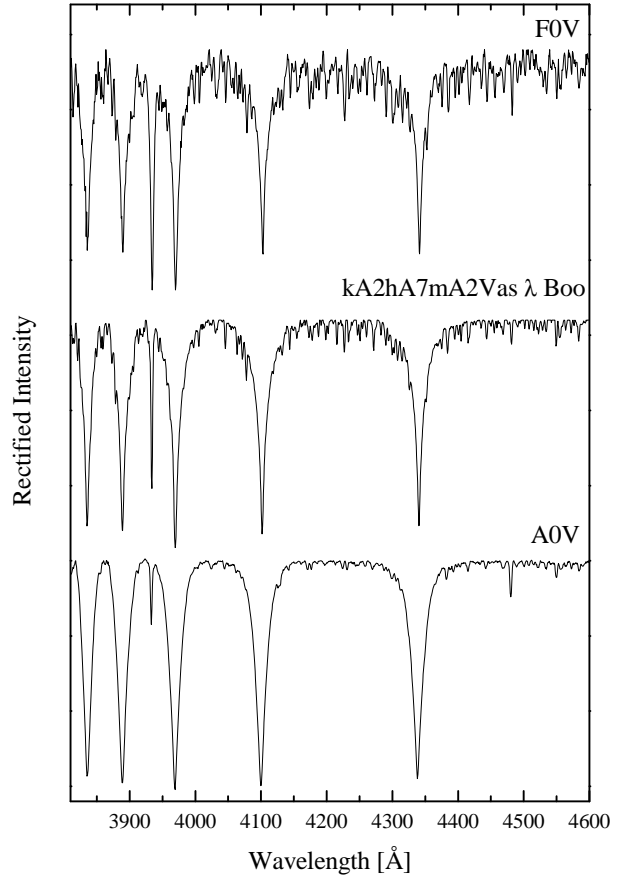}
\caption[]{A new high signal-to-noise classification resolution spectrum of 
HD~210111 collected with Goodman Spectrograph attached to the SOAR telescope 
together with two corresponding MK standards (HR~3314, A0\,V and HD~23585, F0\,V)
taken from \citet{Gra88}}.
\label{class}
\end{center}
\end{figure}

Additional high resolution spectra were obtained at the 
1.5 meter telescope (20./21.10.2009) at the Cerro Tololo 
Inter-American Observatory and the 2.15 meter telescope (05./06.09.2011) 
at the Complejo Astron{\'o}mico El Leoncito (CASLEO) with the Echelle de Banco Simmons (EBASIM) Spectrograph. 
The integration times were set to ten minutes and one hour, respectively. The
spectral resolutions are comparable to that of UVES but the signal-to-noise ratio
is much lower. As supplement, a classification resolution (R\,$\approx$\,500) spectrum (Fig. \ref{class}) 
was obtained at the Southern Astrophysical Research (SOAR) 4.1 meter telescope at Cerro Pachon using the Goodman Spectrograph 
in the night of 26./27.07.2010. 

In a first step, all the spectroscopic data have been reduced using standard IRAF routines.
The standard reduction includes: bias subtraction, flat-fielding, cosmic ray cleaning and wavelength 
calibration.

\section{Target characteristics} \label{target}

HD~210111 (HR~8437, HIP~109306) is one of the southern bright (V\,=\,6.83\,mag)
$\lambda$ Bootis prototype stars. It is classified as kA2hA7mA2\,Vas $\lambda$ Boo 
with peculiar hydrogen lines by \citet{Gra88}. In Fig. \ref{class} we present a new
high signal-to-noise classification resolution spectrum of the target. Together with
the appropriate MK standard stars, it shows the uniqueness of the overall
spectral features.

The high resolution spectrum (R\,$\approx$\,50\,000) presented
by \citet{Hol91} indicate asymmetric line profiles which were not 
discussed in that paper. It is interesting to note that HD~183324 (HR~7400),
another well established member of the $\lambda$ Bootis group, 
shows similar line profile characteristics as HD~210111 according to their presented
spectra. This star might also be an undetected SB2 system.
On the other hand, \citet{Hol95} observed a clean, unperturbed rotationally
broadened \hbox{Ca\,{\sc ii}}~K line. This can be understood within the framework
of a small radial velocity difference between the two binary components, the broadening of the 
\hbox{Ca\,{\sc ii}}~K line
and the spectral resolution. Even in our spectrum, this line (as well as the hydrogen lines)
are not affected by the SB2 characteristics.
In the catalogue of \citet{Gre99} this star is marked as a suspected
binary system. \citet{Far04} suspected that HD~210111 could be an undetected SB system on
the basis of a cross-correlation of three spectra (R\,$\approx$\,28\,000) 
observed in 1993 and 1994.
However, this material was not sufficient to draw any decisive conclusions.

The $\delta$ Scuti type pulsation of this object was detected in 1994 and a 
global observational campaign was reported by \citet{Bre06}. In total, they found
thirteen statistically significant pulsation frequencies with very small photometric 
amplitudes in the visual. It is well known that diffusion and accretion affect the 
pulsation frequencies of stars at the upper main sequence (\citealt{Tur02}).
In $\lambda$ Bootis stars, the opacity in the metal bump will be
significantly lowered. However, only little direct pulsational excitation from 
Fe-peak elements was found, but effects due to settling of helium along with the 
enhancement of hydrogen are important. Nevertheless, the structure of the star
is changed and thus the frequencies of the excited modes. Since the effects are
rather small, new satellite based observations with Convection, Rotation and Planetary Transits (CoRoT) or
Microvariability \& Oscillations of Stars (MOST) are clearly
needed.

A significant IR-excess from IRAS data due to a circumstellar disk
was detected for HD~210111 (\citealt{Pau03}). Since we find that 
both components of this system are
very similar with an equal luminosity (Sect. \ref{redan}), the conclusions
about the characteristic dust temperature, the
fractional dust luminosity, and the radiative equilibrium distance
of the above mentioned paper are still valid.

The overall characteristics of this star makes it an excellent test
case for the published models explaining the $\lambda$ Bootis phenomenon.
With the tools of asteroseismology further insights of the stellar atmospheres
should be possible whereas a detailed analysis of its environment will allow
to understand the source as well as the mechanism of accretion. 

\section{Analysis} \label{redan}

The ATLAS9 model atmospheres, including the treatment of convection by
\citet{Can96}, were calculated with
scaled solar abundances using pretabulated opacity distribution
functions taken from the Vienna New
Model Grid of Stellar Atmospheres (NEMO) (\citealt{Hei02b}). 
All models are scaled to the solar abundance values as listed by \citet{Gre98},
except for C, N and O for which the values of \citet{Asp05} were used.
The microturbulent velocity $\xi$ was set to 2.0\,km\,s$^{-1}$.
This is a typical value used for previous abundances studies 
of $\lambda$ Bootis and A-type stars. Changing $\xi$ to a value
of 3.5\,km\,s$^{-1}$ (see \citealt{Hei02a}) alters the spectral
broadening by an amount which is negligible compared to the
rotational broadening. The atomic transition parameters for the spectrum
synthesis have been taken from the Vienna Atomic Line
Database (VALD; \citealt{Kup99}).

Starting values for an effective temperature, surface gravity and [M/H]
were based on the results by \citet{Stu93}. The continuum flux was calculated
with the program SYNTH (\citealt{Pis92}). In order to
derive a ``composite'' spectrum, the individual spectrum of each
component was shifted by an arbitrary relative radial velocity, weighted
by the continuum flux ratios and added. 

First of all, we estimated the rotational 
velocity $v \sin i$ by inspecting a composite spectrum with several different
values for both components. A value of $v \sin i$\,$\approx$\,30\,km\,s$^{-1}$ for both
components fitted the observations best and was set, accordingly, for the further 
analysis. The heuristic error of the rotational velocity is $\approx$\,5\,km\,s$^{-1}$.
Prior published values of the rotational velocity for HD~210111 range
from 50 to 60\,km\,s$^{-1}$ on the basis of classification to high resolution
spectroscopy (\citealt{Sol01}). Besides the broadening due to the instrumental profile, 
a radial velocity difference of 50\,km\,s$^{-1}$ at 4200\,\AA\, is needed to convolve two
single line profiles with a $v \sin i$\, of 30\,km\,s$^{-1}$ to a ``composite'' one with
60\,km\,s$^{-1}$. This value can be set as upper limit of the orbital total radial velocity
amplitude for all published data. 

As next step, the radial velocities were manually determined and set to $-$23 and
+18\,km\,s$^{-1}$, respectively. Due to the high resolution of the spectrum and the
wide separation of the components, these values are unambiguous with an accuracy better than
1\,km\,s$^{-1}$. The averaged mean radial velocity of $-$5\,km\,s$^{-1}$
is in the same range as the three measurements ($-$5.13$\pm$0.59\,km\,s$^{-1}$) 
published by \citet{Gre99}. These observations were made in a time interval of 434 days. 
Such a rather constant mean radial velocity could point towards a very long orbital
period of this spectroscopic binary system.

The published values for the effective temperature range from 7400 to 7900\,K whereas
the log\,$g$ ones are between 3.75 and 3.90\,dex (\citealt{Pau02}). We
calculated all combinations of stellar atmospheres with 7000\,$<$\,$T_\mathrm{eff}$\,$<$\,8000\,K,
3.7\,$<$\,log\,$g$\,$<$\,4.0\,dex and [M/H]\,=\,[+0.0,$-$0.5,$-$1.0,$-$1.5]. A possible much cooler
component would have been detected by enhanced X-ray fluxes from the ROSAT measurements (\citealt{Hue98}).
The composite spectra were semi-automatically compared to the observed spectrum in the wavelength 
region between 4500 and 5500\,\AA\,including the most prominent unblended metallic lines. 
In addition, the \hbox{Ca\,{\sc ii}}~K, \hbox{Na}~D, and the \hbox{Ca\,{\sc ii}}
lines in the red spectral region were used.
The ten best fitting composite spectra deduced via a SIMPLEX method described by \citet{Gra03} were 
manually inspected and re-fitted.

\citet{Boh99} presented time series spectroscopy of HD~210111 finding non-radial pulsation (NRP)
of this object with a period of 49 minutes and a peak-to-peak amplitude of 2.5\% of the continuum
in the mean-absolute-deviation. The corresponding radial velocity amplitude of such kind of variation
is below 3\,km\,s$^{-1}$ depending on the detected pulsation mode (\citealt{Kis02}). Our reported
absolute radial velocity difference of 41\,km\,s$^{-1}$ is more than one order of magnitude larger than 
that and can not be explained by NRP. 

Finally, we found that a composite spectrum with equal components fitted the observed spectrum
best. In Fig. \ref{hrs} the observed spectrum in the wavelength region from 5425 to 5470\AA\,
together with the difference to the synthetic one is shown. The fit of the line profiles and
depths is better than 1\% which is very satisfactory. Changing the effective temperatures by $\pm$100\,K
and the surface gravities by $\pm$0.05\,dex already significantly (3$\sigma$ of the estimated 
error due to the signal-to-noise ratio) decreases the quality of the fit. 
For the metallicity we can only state that the best fifteen composite spectra always comprise
two components with [M/H]\,=\,$-$1.0\,dex. No combination with at least one solar abundant component
is able to fit the observed line profiles in a satisfactory way. 

Finally we conclude that HD~210111 consists of two similar (within the estimated errors) stars with
$T_\mathrm{eff}$\,=\,7400\,K, log\,$g$\,=\,3.8\,dex, [M/H]\,=\,$-$1.0\,dex and 
$v \sin i$\,=\,30\,km\,s$^{-1}$. This result perfectly matches the strict limitations of the photometric
7-color Geneva, Str{\"o}mgren $uvby$ and $\Delta$a measurements presented in \citet{Stu06}.

Within the usable wavelength range, we mainly find Fe and Si lines as well as a few Mg, Si, Cr and
Ni lines. For these elements, the best-fit spectra with an abundance of  
$-$1.0\,dex compared to the Sun in each of the components agree with the observed
line profiles which is in line with the result by \citet{Stu93}. 

For a more detailed abundance analysis, additional data, especially with a large
separation of the two components are needed.

We fitted a composite synthetic spectrum with the above listed parameters to the H$\beta$ line profiles
of the available high resolution observations. Besides the known inadequateness of fitting the hydrogen
line core correctly, the normalization of the echelle spectra is quite problematic because the line 
spreads over two consecutive orders. The classification resolution spectrum, on the other hand, allows
only a rough determination and check of the parameter space. The fits themselves result in a good agreement.
However, we are not able to use the H$\beta$ line profiles for further improvements via an iterative
method.

From the parallax measurement of the final release of the Hipparcos catalogue (\citealt{Leu07}), 
we derived a distance of 78$\pm$4\,pc for HD~210111. 
With the absolute bolometric magnitude of the Sun
($M_{\rm Bol}$)$_{\odot}$\,=\,4.75\,mag (\citealt{Cay96}) and
the bolometric correction taken from \citet{Dri00}, the luminosity
(log\,L$_{\ast}$/L$_{\odot}$) was calculated.
As the next step, we used the post-MS evolutionary tracks and isochrones
from \citet{Cla95} to estimate the mass and age. The models were calculated with solar abundances.
That is justified because the stellar abundance is restricted to the surface only. 

Using the derived astrophysical parameters, an age of almost 1\,Gyr for this system
and a mass of 1.9\,M$_{\odot}$ for both components is deduced. The errors are about 10\,\%.
This is in line with the ages published by \citet{Ili95} and \citet{Pau02}.

Its brightness and close distance to the Sun makes HD~210111 to an excellent candidate for follow-up
observations in the NIR and IR not only to study its surrounding environment, i.e. the circumstellar disk 
characteristics, but also to understand the accretion process at such evolved ages.

\section*{Acknowledgements}
This work was supported by the financial contributions of the Austrian Agency for International 
Cooperation in Education and Research (WTZ CZ-10/2010 and HR-14/2010).
UH acknowledges support from the Swedish National Space Board. This paper was partially supported by 
PIP0348 by CONICET. Data from the UVES Paranal Observatory Project 
(ESO DDT Program ID 266.D-5655) were used for this paper. This research has made use of the model 
atmosphere grid NEMO and the Vienna Atomic Line Database VALD.
Based on observations obtained at the Southern Astrophysical
Research (SOAR) telescope, which is a joint project of the
Minist{\'e}rio da Ci{\'e}ncia, Tecnologia, e Inova\c{c}{\~a}o (MCTI) da Rep{\'u}blica
federativa do Brasil, the US National Optical Astronomy Observatory
(NOAO), the University of North Carolina at Chapel Hill
(UNC), and Michigan State University (MSU).

\label{lastpage}

\end{document}